\documentstyle[12pt]{article}
\newcommand{\be}{\begin{equation}}
\newcommand{\ee}{\end{equation}}
\newcommand{\bea}{\begin{eqnarray}}
\newcommand{\eea}{\end{eqnarray}}
\newcommand{\bean}{\begin{eqnarray*}}
\newcommand{\eean}{\end{eqnarray*}}
\font\upright=cmu10 scaled\magstep1
\font\sans=cmss12
\newcommand{\ssf}{\sans}
\newcommand{\stroke}{\vrule height8pt width0.4pt depth-0.1pt}
\newcommand{\Z}{\hbox{\upright\rlap{\ssf Z}\kern 2.7pt {\ssf Z}}}
\newcommand{\ZZ}{\Z\hskip -10pt \Z_2}
\newcommand{\C}{{\rlap{\rlap{C}\kern 3.8pt\stroke}\phantom{C}}}
\newcommand{\R}{\hbox{\upright\rlap{I}\kern 1.7pt R}}
\newcommand{\CP}{\C{\upright\rlap{I}\kern 1.7pt P}}

\begin{document}
\pagestyle{plain}

\title{\vskip -70pt
\begin{flushright}
{\normalsize DAMTP 94-78} \\
\end{flushright}
\vskip 20pt
{\bf Skyrme Crystal from a Twisted Instanton on a Four-Torus} \vskip 20pt}

\author{N.S. Manton\thanks{email nsm10@amtp.cam.ac.uk} \\[10pt]
and \\[10pt]
P.M. Sutcliffe\thanks{email p.m.sutcliffe@amtp.cam.ac.uk} \\[30pt]
{\sl Department of Applied Mathematics and Theoretical Physics} \\[5pt]
{\sl University of Cambridge} \\[5pt]
{\sl Silver St., Cambridge CB3 9EW, England} \\[20pt]}
\date{September 1994}
\maketitle

\begin{abstract}
We describe how an approximation to the minimal energy Skyrme crystal
can be obtained from the holonomy of a Yang-Mills instanton.
The appropriate instanton is twisted on a four-torus and has instanton
number equal to one half. It generates a Skyrme field
with the correct topological and symmetry properties of the crystal.
An explicit solution for the instanton is not known, but
an analytical fit to numerical data is available and
using this we obtain a Skyrme crystal whose energy is only 2\% above
that of the (numerically) known solution.

\end{abstract}
\newpage
\section{Introduction}
The Skyrme model is an effective theory of low energy hadron physics
in which nucleons are described by topological solitons. The baryon
number $B$ is identified with the topological charge (or number of solitons)
and it is of interest to determine the classical soliton solutions,
since they describe the classical state of the nucleons. However, the
Skyrme field equation is a nonlinear partial differential
equation which is difficult to solve. In fact, even for the simplest
case of $B=1$ the minimal energy solution is only known numerically.

In a region with high baryon number, the solitons form a crystal.
Klebanov \cite{K} began the
study of Skyrme crystals and over the last 10 years various numerical
and analytical techniques have been applied to find the
solution with minimum energy per baryon.

It has proved useful for understanding the solitons
of the Skyrme model to construct approximate solutions
by calculating the holonomy of Yang-Mills instantons \cite{AMa}.
This method has led to a good understanding of the geometry
of the $B=2$ sector \cite{AMb}, and the
tetrahedral and cubic symmetries of the minimal energy $B=3$ and $B=4$
solutions respectively \cite{LM}. The instanton method has
computational advantages, and it provides a means
by which the full soliton theory may be truncated to a finite number
of degrees of freedom. However, one area in which it has yet to make a
contribution is that of the Skyrme crystal.
In this letter we take the first steps in applying the instanton method
to the Skyrme crystal by identifying a suitable Yang-Mills instanton.
We also make a qualitative calculation based on numerical data for the
instanton.

\section{Skyrme Crystal}
The field $U({\bf x})$ of the static Skyrme model is an $SU(2)$ matrix-valued
function of the space coordinates ${\bf
x}=(x_1,x_2,x_3)=(x,y,z)$. It may be expressed in terms of sigma and
pion fields as
\be
U=\sigma + i\mbox{\boldmath $\pi\cdot\tau$}
\label{pion}
\ee
where $\mbox{\boldmath $ \pi$}=(\pi_1,\pi_2,\pi_3)$
is a triplet of pion fields
and {\boldmath $\tau $} denotes the Pauli matrices. The normalization
constraint is $\sigma^2+\mbox{\boldmath $\pi .\pi$}=1$. For finite
energy fields, it is required that as $|{\bf x}|
\rightarrow \infty$, $U({\bf x}) \rightarrow 1$.
The energy density is
\be {\cal E}=-\frac{1}{24\pi^2}{\rm Tr}(R_iR_i)
-\frac{1}{192\pi^2}{\rm Tr}([R_i,R_j][R_i,R_j])
\label{enden}
\ee
and the baryon density is
\be {\cal B}=-\frac{1}{24\pi^2}\epsilon_{ijk}{\rm Tr}(R_iR_jR_k)
\label{bden}
\ee
where $R_i$ is the right current $\partial_iUU^{-1}$. The spatial integrals of
${\cal E}$ and ${\cal B}$ give the energy $E$ and the baryon number
(topological charge) $B$ of the Skyrme field.
The above units are chosen so that the Fadeev-Bogomolny bound on the
energy $E$ is simply
\be E\geq |B|.
\label{bog}
\ee
In flat space this bound cannot be attained (except by the vacuum
solution $U=1$) and the single Skyrmion ($B=1$) solution has energy
$E=1.23$.

With certain relative orientations well separated Skyrmions
attract, and at high density form a crystal.
Klebanov's original crystal \cite{K} was a simple cubic lattice of Skyrmions
whose symmetries were motivated by the attempt to rotate the
Skyrmions so as to maximize the attraction between nearest neighbours.
It has energy per baryon of $E=1.08$.
Other symmetries were
proposed which lead to slightly lower, but not minimal energy crystals
\cite{GMJV}. Following the work of Castillejo et al. \cite{CJJVJ}, and
of Kugler and Shtrikman \cite{KS}, it is now understood
that it is best to arrange the
Skyrmions initially as a face-centred cubic lattice.
Their orientations can be
chosen very symmetrically to give maximal attraction between all nearest
neighbours. This configuration relaxes to the minimal
energy crystal, the Skyrme crystal, and in the process the symmetry
increases further. The numerical investigations of
\cite{CJJVJ}, and the variational approach of \cite{KS} using fourier series,
lead to the conclusion that the Skyrme crystal
has energy per baryon of $E=1.038$
and a cubic unit cell of side $L=4.7$.
It is a crystal of half-Skyrmions, as a unit cell
contains half a unit of baryon number. The fields are
strictly periodic after translation by $2L$ in the $x$,$y$ or $z$
directions, and within a cube of side $2L$ the baryon number is $B=4$.
The crystal symmetries, which differ from those of the
crystals proposed in refs. [1] and [5], involve translations,
reflections and rotations
of the space coordinates combined with O(4) rotations of the sigma and
pion fields. Explicitly, the symmetry generators are \cite{KS}
\bea
(x,y,z)\rightarrow&(-x,y,z)\hskip 40pt \& \
&(\sigma,\pi_1,\pi_2,\pi_3)\rightarrow(\sigma,-\pi_1,\pi_2,\pi_3) \label{sa}\\
(x,y,z)\rightarrow&(y,z,x)\hskip 50pt \& \
&(\sigma,\pi_1,\pi_2,\pi_3)\rightarrow(\sigma,\pi_2,\pi_3,\pi_1) \label{sb}\\
(x,y,z)\rightarrow&(x,z,-y)\hskip 40pt \& \
&(\sigma,\pi_1,\pi_2,\pi_3)\rightarrow(\sigma,\pi_1,\pi_3,-\pi_2) \label{sc}\\
(x,y,z)\rightarrow&(x+L,y,z) \hskip 25pt \& \
&(\sigma,\pi_1,\pi_2,\pi_3)\rightarrow(-\sigma,-\pi_1,\pi_2,\pi_3).
\label{sd}
\eea

The fields obtained numerically, or by optimising the fourier series, are
very well approximated by the analytical formulae \cite{CJJVJ}
\begin{eqnarray}
\sigma&=&c_1c_2c_3 \nonumber\\
\pi_1&=&-s_1\sqrt{1-\frac{1}{2}s_2^2-\frac{1}{2}s_3^2+\frac{1}{3}s_2^2s_3^2}
\hskip 20pt \mbox{and cyclic} \label{jack}\end{eqnarray}
where
\be
s_i=\sin(\frac{\pi x_i}{L}) \hskip 10pt \mbox{and} \hskip 10pt
c_i=\cos(\frac{\pi x_i}{L}).
\ee

\section{Twisted Instantons on $T^4 \hskip -18.5pt T^4$}
The pioneering work on Yang-Mills instantons on a four-torus $T^4$ with
twisted boundary conditions was done by 't Hooft \cite{tH},
motivated by the study of quark confinement in QCD. Let $A_\mu$ be the
$su(2)$-valued Yang-Mills gauge potential, with field strength
$F_{\mu\nu}$. Here greek indices take the values 0 to 3, and latin
indices 1 to 3, and we write $x_\mu=(t,x,y,z)$, with $t$ the euclidean
time coordinate. We model the four-torus by
the euclidean box defined by $0\leq x_\mu\leq a_\mu$.
To give the  boundary conditions we introduce $SU(2)$-valued twist
matrices $\Omega_\mu$ each of which is independent of the $\mu$th
spacetime coordinate. The twisted boundary conditions are that the
gauge potentials are periodic modulo gauge transformations
\be
A_\nu(x_\mu=a_\mu)=\Omega_\mu^{-1} A_\nu(x_\mu=0)\Omega_\mu
+\Omega_\mu^{-1}\partial_\nu\Omega_\mu.
\label{tbc}
\ee
The notation $A_\nu(x_\mu=a_\mu)$
means that the $\mu$th coordinate is set equal to $a_\mu$ but
the remaining coordinates are arbitrary.
On 2-faces of the box the compatibility conditions which
arise from (\ref{tbc}) are
\be
\Omega_\mu(x_\nu=a_\nu)\Omega_\nu(x_\mu=a_\mu)=
\Omega_\nu(x_\mu=a_\mu)\Omega_\mu(x_\nu=a_\nu)Z_{\mu\nu}
\label{cc}
\ee
where each $Z_{\mu\nu}$ is an element of the centre $\ZZ$ of $SU(2)$.
We introduce the gauge invariant antisymmetric twist integers
$n_{\mu\nu}\in\Z \ \mbox{(mod 2)}$ by
\be
Z_{\mu\nu}=\exp(i\pi n_{\mu\nu}).
\ee
It is these that are the important quantities, since
any twist matrices with the same set of twist integers are gauge
equivalent. It is convenient to collect the twist integers into a
magnetic flux vector
${\bf m}$ and an electric flux vector ${\bf k}$, where
$n_{ij}=\epsilon_{ijk}m_k$ and $k_i=n_{0i}$.

't Hooft pointed out that the usual expression for the instanton
number does not have to be integer valued in the presence of twist.
In fact
\be
N=\frac{1}{16\pi^2}\int_{T^4}{\rm Tr}(\ast F_{\mu\nu}F^{\mu\nu}) d^4x
=q+\frac{{\bf k\cdot m}}{2}
\label{inum}
\ee
where $q$ is an integer. The mathematical explanation of this result
can be found in \cite{PvB}, \cite{CN}. The expression (\ref{inum}) is minus the
second Chern number for an $SU(2)$ gauge theory, but since the gauge
functions are only defined modulo the centre of the gauge group, we
are in fact dealing with the gauge group $SU(2)/\ZZ = SO(3)$, and that
is why $N$ can be an integer or half-integer.

Many elegant mathematical results are known for twisted instantons
on $T^4$, such as an existence proof \cite{Sed}, a (non-explicit)
construction of the appropriate bundles in terms of K-theory \cite{CN}
and a Nahm duality transformation \cite{PBPvB}. The moduli (parameter)
space of instantons of instanton number $N$ is $8N$, that is, an integer
multiple of four \cite{DK}.

\section{Skyrme Crystal from an Instanton}
We now show how an approximation to the Skyrme crystal
described in section 2 can be obtained from the holonomy of a suitable
instanton on $T^4$.

In order to obtain a Skyrme field with the correct cubic symmetry we
set $a_i=L$ and $a_0=T$, and choose
isotropic magnetic and electric flux vectors ${\bf m}=(1,1,1)$ and
${\bf k}=(1,1,1)$. Then (\ref{inum}) becomes $N=q+\frac{3}{2}$.
There are explicit abelian solutions \cite{tH} for $N=\frac{3}{2}$,
when the four-torus has the same length in each direction, but we
are interested in approximating a crystal composed of half-Skyrmions
so we wish to consider the case $N=\frac{1}{2}$. Note that the
Yang-Mills action on $T^4$ is bounded by $8\pi^2|N|$, so the
$N=\frac{1}{2}$ instantons are the minimal action configurations
outside the vacuum sector. Moreover, their moduli space is
four-dimensional and simply parametrises the translates, and certain
discrete gauge transformations of, a unique $N=\frac{1}{2}$ instanton
on the given torus. This instanton is invariant under
the cubic subgroup of the spatial rotation group, and may be assumed
to be centred with its maximum action density at $x=y=z=L/2$ and $t=T/2$.

By choosing explicit twist matrices we perform a partial gauge
fixing. Here it is convenient to use the Pauli matrices for the
spatial twist matrices
\be
\Omega_j=i\tau_j.
\label{pauli}
\ee
It is a simple matter to check that this choice satisfies
the spatial components of
(\ref{cc}) with the twist integers given earlier.
It is also easy to construct a twist matrix $\Omega_0$
which satisfies the remaining components of (\ref{cc}). Although
we shall not need an explicit form for this matrix we give a possible
choice for completeness,
\be
\Omega_0=(\prod_{i=1}^3 c_i)+i\sqrt{\frac{1-(\prod_{i=1}^3 c_i)^2}
{{\bf c\cdot c}}}\mbox{\boldmath {\bf c}$\cdot\tau$}
\label{omo}
\ee
where $c_i=\cos({\pi x_i}/{L})$.\\

The Skyrme field generated by the instanton is, by definition, the
holonomy of the instanton along all lines parallel
to the euclidean time axis, {\it ie}
\be
U({\bf x})=\left({\cal P}\exp\int_0^T A_0(x_0,{\bf x})\
dx_0\right)\Omega_0({\bf x})
\label{hol}
\ee
where ${\cal P}$ denotes path ordering. Note the inclusion of the
twist matrix since the path must be a closed loop. The Skyrme field is
defined on a cube of side $L$, and is extended by symmetry and
continuity to a crystal with this cube as unit cell.
The usual argument for instantons on $\R^4$ can be
repeated in this case to show that the instanton number $N$ equals
the baryon number $B$ of the Skyrme field that it generates. Hence the
Skyrme field has $B=\frac{1}{2}$ in a unit cell.

We can show,
without using the explicit form (\ref{omo}), that the Skyrme
field has the symmetry (\ref{sd}) of a half-Skyrmion crystal.
Using (\ref{tbc}) and (\ref{pauli}) we have
\be
U(x+L,y,z)=\tau_1\left({\cal P}\exp\int_0^T A_0(x_0,x,y,z)\
dx_0\right)\tau_1 \ \Omega_0(x+L,y,z).
\label{shift}
\ee
Equation (\ref{cc}) implies that
\be
\Omega_0(x+L,y,z)=-\tau_1\Omega_0(x,y,z)\tau_1,
\label{shifto}
\ee
and combining (\ref{shift}) and (\ref{shifto}) gives
\be
U(x+L,y,z)=-\tau_1U(x,y,z)\tau_1,
\ee
which is equivalent to symmetry (\ref{sd}). The symmetries
(\ref{sa})-(\ref{sc}) are a consequence of the cubic symmetry of the
instanton.

To progress, we would like to have an
explicit analytical solution for the instanton of interest, but this is
not available. Fortunately some numerical studies have been performed
by Garc\'ia P\'erez et al. \cite{GPGAS} and
we can make use of their
results as a first step in determining the instanton-generated
Skyrme crystal, and estimating how good an approximation it
is. Garc\'ia P\'erez et al. treat the Yang-Mills theory as a
lattice gauge theory and search for the minimal action configuration
using the lattice cooling method. For the case $T=\infty$ they
provide an analytical fit to the numerical data for the holonomy, and
it is this result
which we shall use here. Although $T=\infty$
the instanton is still localized in the euclidean time direction, with
a scale which is determined by the spatial length of the four-torus, $L$.
(In this limit, one lets the range of $t$ be
$-\infty < t< \infty$ and assumes that the instanton has its
maximal action density at $t=0$.)
The analytical fit from \cite{GPGAS} for the holonomy in the euclidean
time direction is
\be U({\bf x})=
(\prod_{i=1}^3 h_i)+i\sqrt{\frac{1-(\prod_{i=1}^3 h_i)^2}
{{\bf f\cdot f}}}\mbox{\boldmath {\bf f}$\cdot\tau$}
\label{ah}
\ee
where
\bea
h_i=\cos(\frac{\pi x_i}{L})(1+\alpha\sin^2(\frac{\pi x_i}{L})), \\
f_i=-\sin(\frac{\pi x_i}{L})(1+\beta\cos^2(\frac{\pi x_i}{L})).
\label{mandf}
\eea
Here $\alpha=-0.196$ and $\beta=-0.172$ are constants whose values are
determined by the fit.
Note that this Skyrme field has a similar form to the approximation
to the Skyrme crystal (\ref{jack}) mentioned earlier.
It is a simple
matter to check explicitly that it has all the
symmetry properties (\ref{sa}) to (\ref{sd}) of the Skyrme crystal.

It is relatively easy to obtain highly accurate
numerical integrals for quantities such as the baryon number and
energy of the Skyrme field (\ref{ah}) since the
differentiation of $U$ can be performed analytically, and furthermore
the region of integration is finite.
Integrating the baryon density (\ref{bden}) over a cube
of side $L$ confirms the baryon number to be $B=\frac{1}{2}$,
and we find the energy per baryon is
\be
E=0.1069L+\frac{2.6146}{L}.
\ee
Hence the minimum energy is $E=1.058$ at $L=4.9$,
with both these figures being reasonably close to the Skyrme crystal values.
One may take the view that the crystal size should be
fixed at the value $L=4.7$.
The approximation (\ref{ah}) then contains no free parameters
at the expense of an energy increase of 0.1\%.
The fact that the energy is only 2\% above that of the Skyrme crystal is
encouraging, given that we are using only a fit
to numerical data. Errors will be introduced both in the numerical
construction of the instanton and in the analytical fitting process,
and it would be very interesting if a more accurate
numerical method could be employed in
calculating the instanton and its holonomy.

In \cite{GPGAS} it is noted that moderate variations of the parameters
$\alpha$ and $\beta$ may not significantly alter the quality of the fit to the
numerical data. To see whether a more accurate
scheme  may produce a better approximate
Skyrme crystal, we allow $\alpha$ and
$\beta$ to be arbitrary parameters. Using a steepest decent algorithm
in $(\alpha,\beta)$ space we find that the minimum crystal energy
is obtained when $\alpha=-0.02$ and $\beta=-0.25$. At these values
the energy is $E=1.040$ for a crystal size $L=4.7$, {\it ie} the energy
is within $\frac{1}{5}$\% of the true minimum.

Allowing $T$ to vary could produce a better approximation to the
Skyrme crystal. In \cite{EK} it was shown that a
slightly better approximation to the $B=1$ Skyrmion in \R$^3$ can be
obtained from an instanton on \R$^3$ times a finite euclidean time
interval (a caloron) than from an instanton on \R$^4$. Moreover,
a lower dimensional
analogue of the procedure described  in this letter exists \cite{PMS},
where sine-Gordon kink chains are approximated by the holonomies
of \CP$^1$ instantons on $T^2$. In that example a much
better approximation is obtained by having a finite length for the
torus in the euclidean time direction. If this analogue result is a
good guide to what happens in the case of the Skyrme crystal then one must
certainly consider finite $T$.

\section{Conclusion}
In this letter we have identified the Yang-Mills instanton whose holonomy
produces a Skyrme field which has the topological and symmetry
properties of the Skyrme crystal. Since an explicit solution is not
known for the instanton it is not an easy task to obtain accurate quantitative
information about the generated crystal. However,
by using numerical work of Garc\'ia P\'erez et al. \cite{GPGAS}
we have obtained some initial results
which are encouraging. We hope that a more detailed investigation of
this problem will prove fruitful.\\

\noindent{\bf Acknowledgements}\\
NSM is grateful to Charles Nash for very helpful discussions and
correspondence. PMS would like to thank the EPSRC for a research
fellowship.
This work was supported in part by PPARC.

\end{document}